\documentclass[preprint,12pt]{elsarticle}
\usepackage{amssymb}
\usepackage{graphicx}
\usepackage{dcolumn}
\usepackage{bm}
\usepackage{amsmath}
\usepackage{latexsym}
\usepackage{epsfig}
\usepackage{amsbsy}
\usepackage{array}
\usepackage{setspace}
\usepackage{bm}

\begin{document}

\begin{frontmatter}



\title{The propagation of a polarized
Gaussian beam in a smoothly inhomogeneous isotropic medium}


\author[label1]{Hehe Li\corref{cor1}}
\author[label1,label2]{Peiyong Ji}
\address[label1]{Department of Physics,
Shanghai University, Shanghai 200444, China}
\address[label2]{The Shanghai Key Laboratory of
Astrophysics, Shanghai 200234, China} \cortext[cor1]{Email:
$hehe_{-}li@shu.edu.cn$}

\begin{abstract}
We present a description of the evolution of a polarized Gaussian
beam in a smoothly inhomogeneous isotropic medium in frame of the
eikonal-based complex geometrical optics which describes the phase
front and the cross section of the Gaussian beam using the quadratic
expansion of the complex-valued eikonal. The linear complex-valued
eikonal components are introduced to describe the influence of the
spin-orbit interaction and the deformation of a polarized Gaussian
beam on the propagation firstly in this paper. In an inhomogeneous
medium, the interaction between the polarization and the rotation
deformation of the light beam is presented besides the spin-orbit
interaction, it corresponds to the spin-intrinsic orbital angular
momentum interaction and makes the correction for the spin Hall
effect of a polarized Gaussian light beam.
\end{abstract}

\end{frontmatter}


\section{Introduction}
The propagation of a polarized light has been investigated in
various physical environment[1-16]. In the process of the
propagation, there are two important observable effects, the spin
Hall effect[1-12] and the Berry phase[17]£¬ which describe the
splitting of the propagation trajectory of different polarized light
and the rotation of the polarization ellipse[18,19,20] respectively.
The two phenomena are caused by the spin-orbit interaction of
photons, which describes the interaction between the polarization
and the extrinsic orbital angular momentum of light[1,4,10].

The eikonal-based complex geometrical optics[21,22,23,24] and the
solution of a parabolic-type wave equation have given the
description of the Gaussian beam propagation in an inhomogeneous
medium, and the evolution of the Gaussian beam reduces to the
solution of the Riccati type ordinary differential equations. One
knows that the quadratic expansion complex-valued eikonal[22,23] is
introduced to describe the phase front and the cross section of the
Gaussian beam in frame of the eikonal-based complex geometrical
optics. There is the deformation of the Gaussian which is due to the
medium inhomogeneity[22,23]. we have known there is the spin-orbit
interaction for a polarized light beam, but whether there is the
interaction between the polarization and the deformation of the
light beam? In this paper, this problem is investigated using the
eikonal-based complex geometrical optics.

In frame of the eikonal-based complex geometrical optics, One knows
that there are no the linear components in the complex-valued
eikonal[23] when the Gaussian beam propagates in the homogeneous
isotropic medium, but the inhomogeneous isotropic medium can be
considered as the weakly anisotropic medium for different polarized
light in the first-order geometrical optics approximation[3]. In
order to describe the perturbation in the propagation, the linear
complex-valued eikonal components are introduced in this paper. We
obtain the result that the propagation of a polarized Gaussian beam
in an inhomogeneous isotropic medium is affected by the spin-orbit
interaction and the rotation deformation of the beam. The spin Hall
effect of a polarized light beam[25] describes the transverse shift
of the beam as a whole and is determined by the spin-orbit
interaction. The rotation deformation of the light beam produces an
equivalent intrinsic orbital angular momentum[26,27,28], there must
be the interaction between the polarization and the rotation
deformation of the light beam. This interaction makes the correction
for the spin Hall effect of a polarized light beam.

The paper is organized as follows. First, the eikonal equation and
amplitude transport equation which include the first-order
geometrical optics approximation correction are derived. Then, the
linear complex-valued eikonal components are introduced to describe
the influence of the medium's inhomogeneity on the propagation of a
polarized beam within the paraxial approximation. And the
differential equations describing the evolution of a polarized
Gaussian beam are obtained. Finally, we reexamine the spin Hall
effect of a polarized Gaussian beam and obtain result that both the
spin-orbit interaction and the deformation of a polarized Gaussian
affect the propagation trajectory of the light beam.
\section{The eikonal equation in first-order geometrical optics approximation}
Let's consider the propagation of a monochromatic linearly polarized
light in a smoothly inhomogeneous isotropic medium with dielectric
permittivity $\varepsilon(\textbf{r})$. There is a small geometrical
optics parameter $\mu_{GO}$[21,29],
\begin{equation}
\mu_{GO}=\frac{\lambda}{L}\ll 1,
\end{equation}
where $L\sim |\nabla\varepsilon/\varepsilon|^{-1}$ is the
characteristic scale of the medium inhomogeneity,
$\lambda=2\pi/k_{0}=2\pi c/\omega$ is the wavelength in vacuum,
$\omega$ is the angular frequency. Next, the eikonal equation will
be derived in the first-order approximation in $\mu_{GO}$.

Maxwell equations have the form
\begin{equation}
\nabla\times(\nabla\times\textbf{E})+\frac{\varepsilon}{c^{2}}\frac{\partial^{2}\textbf{E}}{\partial
t^{2}} =0,
\end{equation}
then
\begin{equation}
\nabla^{2}\textbf{E}-\frac{\varepsilon}{c^{2}}\frac{\partial^{2}\textbf{E}}{\partial
t^{2}}-\nabla(\nabla\cdot\textbf{E})=0.
\end{equation}
The last term in the left-hand side of Eq.(3) corresponds to the
spin-orbit interaction of electromagnetic waves[1,10]. In
geometrical optics, the propagation of the electromagnetic wave can
be described by the ray trajectory in the short-wavelength limit. A
ray-accompanying frame $(\eta_{1},\eta_{2},\tau)$ is introduced with
the unit vectors $(\textbf{e}_{1},\textbf{e}_{2},\textbf{l})$,
$\tau$ is the parameter along the ray, which is connected with the
ray arc length $s$ by the relation $d\tau=d
s/\sqrt{\varepsilon_{c}}$.
$\varepsilon_{c}=\varepsilon(\textbf{r}_{c})$,
$\textbf{r}_{c}=(0,0,\tau)$ is the radius vector for the central ray
in zero-order approximation in $\mu_{GO}$. The Lam$\acute{e}$
coefficients of the coordinate frame $(\eta_{1},\eta_{2},\tau)$
are[30,31]
\begin{equation}
h_{1}=h_{2}=1,~~~~~
h=h_{\tau}=\sqrt{\varepsilon}\left[1-\frac{\mathbf{\eta}\cdot\nabla_{\perp}\varepsilon}{2\varepsilon}\right]\mid_{\textbf{r}=\textbf{r}_{c}},
\end{equation}
where
\begin{equation}
\nabla=\textbf{e}_{1}\frac{\partial}{\partial\eta_{1}}+\textbf{e}_{2}\frac{\partial}{\partial\eta_{2}}
+\textbf{l}\frac{1}{h}\frac{\partial}{\partial\tau},~~~~~
\nabla_{\perp}=\textbf{e}_{1}\frac{\partial}{\partial\eta_{1}}+\textbf{e}_{2}\frac{\partial}{\partial\eta_{2}}.
\end{equation}
The electric field of the electromagnetic wave in the ray
coordinates is
$\textbf{E}=E_{1}\textbf{e}_{1}+E_{2}\textbf{e}_{2}+E_{\parallel}\textbf{l}=\textbf{E}_{\perp}+E_{\parallel}\textbf{l}$,
$\textbf{E}_{\parallel}\ll\textbf{E}_{\perp}$. Keeping the terms up
to the $\mu_{GO}$, the wave equation for the transverse electric
field $\textbf{E}_{\perp}$ can be written in the form
\begin{equation}
\nabla^{2}\textbf{E}_{\perp}+k_{0}^{2}\varepsilon\textbf{E}_{\perp}+(\nabla\ln
\varepsilon\times\nabla)\times\textbf{E}_{\perp}=0,
\end{equation}
the third term in the left-hand side of Eq.(6) corresponds to the
spin-orbit interaction, it originates from the third term in the
left-hand side of Eq.(3). One knows a linearly polarized
electromagnetic wave is a superposition of right-hand and left-hand
circularly polarized waves,
$\textbf{E}_{\perp}=E_{+}\textbf{e}_{+}+E_{-}\textbf{e}_{-}$,
\begin{equation}
\textbf{e}^{\pm}=\frac{1}{\sqrt{2}}(\textbf{e}_{1}\pm
i\textbf{e}_{2}),~~~~~ E^{\pm}=\frac{1}{\sqrt{2}}(E_{1}\mp iE_{2}).
\end{equation}
Because of
\begin{equation}
\textbf{l}\times\textbf{e}^{\sigma}=-i\sigma\textbf{e}^{\sigma},
\end{equation}
we obtain the wave equation from Eq.(6),
\begin{equation}
k_{0}^{-2}\nabla^{2}E^{\sigma}+\varepsilon
E^{\sigma}-ik_{0}^{-2}\sigma\textbf{l}\cdot(\nabla\ln
\varepsilon\times\nabla)E^{\sigma}=0,
\end{equation}
where $\sigma=\pm1$ denote the wave helicity of right and left
circular polarizations. One can takes $E^{\sigma}$ as wave function
of photons and introduces the momentum operator of photons
$-ik_{0}^{-1}\nabla$, Eq.(9) can be considered as the
schrodinger-type equation of photons, the Hamiltonian of
photons[1,4,10] which include the spin-orbit interaction is
obtained. Then the equation of motion for photons[1-4,6,10] can be
derived using the canonical equations, which describes the spin Hall
effect of photons.

The eikonal equation which includes the spin-orbit interaction also
can be derived from the wave equation (9). From Eq.(7), $E^{\sigma}$
can be written in the following form
\begin{equation}
E^{\sigma}=A(\textbf{r})\exp[ik_{0}\psi(\textbf{r})]\exp(-i\sigma\frac{\pi}{4}).
\end{equation}
Substituting Eq.(10) into Eq.(9), the eikonal equation and amplitude
transport equation are obtained in the first-order geometrical
optics approximation,
\begin{equation}
(\nabla\psi)^{2}=\varepsilon+\sigma\left(\frac{\nabla\ln\varepsilon}{k_{0}}\times\nabla\psi\right)\cdot\textbf{l},
\end{equation}
\begin{equation}
2\nabla\psi\cdot\nabla
A+A\nabla^{2}\psi=\sigma\left(\frac{\nabla\ln\varepsilon}{k_{0}}\times\nabla
A\right)\cdot\textbf{l}.
\end{equation}
The second term in the right-hand side of Eq.(11) and Eq.(12) are
the spin-orbit correction which corresponds to the first-order
approximation in $\mu_{GO}$. The eikonal equation (11) is same as
the equation of motion derived from the photon's Hamiltonian, and it
also describes the spin Hall effect of photons. The amplitude
transport equation (12) describes the variation of the amplitude
along the ray. If we take $A(\textbf{r})=A(\tau)$, the right-hand
side of Eq.(12) disappear, then $\nabla\cdot(A^{2}\nabla\psi)=0$,
this form means the conservation of the energy flux along the ray.
Next, the propagation of a polarized Gaussian beam is investigated
based on Eq.(11) in the paraxial approximation.
\section{The evolution of a polarized Gaussian beam in an inhomogeneous isotropic medium}
In order to describe the propagation of the Gaussian beam in the
paraxial approximation, two small parameters are introduced,
\begin{equation}
\mu_{1}=\frac{\lambda}{w},~~~~~ \mu_{2}=\frac{w}{L},
\end{equation}
where $w$ is the characteristic beam width. $\mu_{1}$ and $\mu_{2}$
describe the angle of the beam diffraction widening and the
influence of the medium inhomogeneity on the diffraction
respectively[23]. Three parameters, $\mu_{1}$ , $\mu_{2}$ and
$\mu_{GO}$ follow relation
\begin{equation}
\mu_{GO}\ll\mu_{1},\mu_{2}.
\end{equation}
The propagation of the Gaussian beam in an inhomogeneous medium has
been investigated with an accuracy of $\mu_{1}^{2}$ and
$\mu_{2}^{2}$[23,30,31], we will keep the order in
$\mu_{GO}\mu_{1}^{2}$ and $\mu_{GO}\mu_{2}^{2}$ because of the
spin-orbit interaction.

In the paraxial approximation, the eikonal can be written as
\begin{equation}
\psi=\psi_{c}(\tau)+\delta\psi(\eta_{1},\eta_{2},\tau),
\end{equation}
where $\psi_{c}(\tau)$ is the eikonal on the central ray and
$\delta\psi$ is a small deviation in the paraxial approximation.
$\delta\psi$ can be written as the following form[23]
\begin{equation}
\delta\psi(\eta_{1},\eta_{2},\tau)=\frac{\eta_{1}^{2}}{2}S_{11}(\tau)+\frac{\eta_{2}^{2}}{2}S_{22}(\tau)+\eta_{1}\eta_{2}S_{12}(\tau)
+\eta_{1}S_{1}(\tau)+\eta_{2}S_{2}(\tau)+S_{0}(\tau),
\end{equation}
where $S_{ij}$ ($i=1,2$ and $S_{12}=S_{21}$) are the quadratic
complex-valued term in $\eta_{i}\eta_{j}$. The real and imaginary
components of $S_{ij}$ describe the shape of the phase front and the
cross section of a polarized Gaussian beam[21-23]. The linear
complex-valued term, $S_{i}$, describe the deviation of the center
of the beam. One knows that the deviation of the center of the beam
is due to the spin-orbit interaction in the first-order geometrical
optics approximation. $S_{0}(\tau)$ is a small additional phase in
the paraxial approximation. Substituting Eq.(15) and Eq.(16) into
the eikonal equation (11),
\begin{equation}
\left(\frac{\partial\psi}{\partial\eta_{1}}\right)^{2}+\left(\frac{\partial\psi}{\partial\eta_{2}}\right)^{2}
+\frac{1}{h^{2}}\left(\frac{\partial\psi}{\partial\tau}\right)^{2}
=\varepsilon+\frac{\sigma}{k_{0}}\left(\frac{\partial\ln\varepsilon}{\partial\eta_{1}}\frac{\partial\psi}{\partial\eta_{2}}
-\frac{\partial\ln\varepsilon}{\partial\eta_{2}}\frac{\partial\psi}{\partial\eta_{1}}\right).
\end{equation}
The dielectric permittivity $\varepsilon$ can be expanded in a
Taylor series,
\begin{equation}
\varepsilon=\varepsilon_{c}+\bm{\eta}\cdot\nabla_{\perp}\varepsilon_{c}+\frac{1}{2}(\bm{\eta}\cdot\nabla_{\perp})^{2}\varepsilon_{c},
\end{equation}
where $\bm{\eta}=\eta_{1}\textbf{e}_{1}+\eta_{2}\textbf{e}_{2}$,
$\nabla_{\perp}=\partial/\partial\bm{\eta}$,
$\nabla_{\perp}\varepsilon_{c}\equiv(\nabla_{\perp}\varepsilon)|_{\textbf{r}=\textbf{r}_{c}}$.
From Eq.(17) and Eq.(18), the eikonal equation has the form
\begin{equation}
h^{2}\left[\left(\frac{\partial\psi}{\partial\eta_{1}}\right)^{2}+\left(\frac{\partial\psi}{\partial\eta_{2}}\right)^{2}\right]
+\left[\frac{\partial\psi_{c}}{\partial\tau}+\frac{\partial\delta\psi}{\partial\tau}\right]^{2}
=h^{2}\varepsilon+h^{2}\frac{\sigma}{k_{0}}\left(\frac{\partial\ln\varepsilon_{c}}{\partial\eta_{1}}\frac{\partial\psi}{\partial\eta_{2}}
-\frac{\partial\ln\varepsilon_{c}}{\partial\eta_{2}}\frac{\partial\psi}{\partial\eta_{1}}\right).
\end{equation}
In zero-order approximation , the ray equation is obtained
\begin{equation}
\frac{\partial\psi_{c}}{\partial\tau}=\varepsilon_{c},
\end{equation}
This equation describes the evolution of the trajectory of the
center of the beam without the influence of the spin-orbit
interaction[29,32].

From Eq.(19) and Eq.(20), we have
\begin{eqnarray}
h^{2}\left[\left(\frac{\partial\psi}{\partial\eta_{1}}\right)^{2}+\left(\frac{\partial\psi}{\partial\eta_{2}}\right)^{2}\right]
+\left[2\frac{\partial\psi_{c}}{\partial\tau}\frac{\partial\delta\psi}{\partial\tau}+\left(\frac{\partial\delta\psi}{\partial\tau}\right)^{2}\right]
&=&\frac{\varepsilon_{c}}{2}(\bm{\eta}\cdot\nabla_{\perp})^{2}\varepsilon_{c}-\frac{3}{4}(\bm{\eta}\cdot\nabla_{\perp}\varepsilon_{c})^{2}\nonumber \\
& &
+h^{2}\frac{\sigma}{k_{0}}\left(\frac{\partial\ln\varepsilon_{c}}{\partial\eta_{1}}\frac{\partial\psi}{\partial\eta_{2}}
-\frac{\partial\ln\varepsilon_{c}}{\partial\eta_{2}}\frac{\partial\psi}{\partial\eta_{1}}\right).
\end{eqnarray}
Eq.(21) can be expanded in $\mathbf{\eta}$, the differential
equations for the complex parameters $S_{ij}$ and $S_{i}$ are
obtained keeping up to the  $\mu_{GO}\mu_{1}^{2}$ and
$\mu_{GO}\mu_{2}^{2}$,
\begin{equation}
\frac{\partial
S_{11}}{\partial\tau}+(S_{11}^{2}+S_{12}^{2})=\alpha_{11},
\end{equation}
\begin{equation}
\frac{\partial
S_{22}}{\partial\tau}+(S_{12}^{2}+S_{22}^{2})=\alpha_{22},
\end{equation}
\begin{equation}
\frac{\partial
S_{12}}{\partial\tau}+S_{12}(S_{11}+S_{22})=\alpha_{12},
\end{equation}
\begin{equation}
\frac{\partial
S_{1}}{\partial\tau}+(S_{1}S_{11}+S_{2}S_{12})-\frac{\sigma}{2k_{0}}\left(\frac{\partial\ln\varepsilon_{c}}{\partial\eta_{1}}S_{12}-
\frac{\partial\ln\varepsilon_{c}}{\partial\eta_{2}}S_{11}\right)=0,
\end{equation}
\begin{equation}
\frac{\partial
S_{2}}{\partial\tau}+(S_{1}S_{12}+S_{2}S_{22})-\frac{\sigma}{2k_{0}}\left(\frac{\partial\ln\varepsilon_{c}}{\partial\eta_{1}}S_{22}-
\frac{\partial\ln\varepsilon_{c}}{\partial\eta_{2}}S_{12}\right)=0,
\end{equation}
\begin{equation}
\frac{\partial S_{0}}{\partial\tau}+\frac{1}{2}(S_{1}^{2}+S_{2}^{2})
-\frac{\sigma}{2k_{0}}\left(\frac{\partial\ln\varepsilon_{c}}{\partial\eta_{1}}S_{2}-
\frac{\partial\ln\varepsilon_{c}}{\partial\eta_{2}}S_{1}\right)=0,
\end{equation}
where
\begin{equation}
\alpha_{ij}=\frac{1}{2}\frac{\partial^{2}\varepsilon_{c}}{\partial\eta_{i}\partial\eta_{j}}
-\frac{3}{4\varepsilon_{c}}\frac{\partial\varepsilon_{c}}{\partial\eta_{i}}\frac{\partial\varepsilon_{c}}{\partial\eta_{j}}.
\end{equation}
Eqs.(22)-(27) describe the evolution of a polarized Gaussian beam in
the smoothly inhomogeneous isotropic medium, the parameters
$\alpha_{ij}$ describe the influence of the medium inhomogeneity on
the propagation of the beam. These equations can be divided into
three groups. The first group is Eqs.(22)-(24) which describe the
evolution of the shape of the phase front and the cross section of a
polarized Gaussian beam[23]. This group of equations has been
obtained in [23] and is independent of the other equations. The
second group is Eq.(25) and Eq.(26) which describe the transverse
shift of the center of a polarized Gaussian beam. The third group is
Eq.(27) which describes the evolution of the additional phase due to
the deformation of the beam.

We have known that Eqs.(22)-(24) keep the terms in $\mu_{i}^{2}$ and
Eqs.(25)-(26) in $\mu_{GO}\mu_{i}^{2}$. The spin Hall effect of a
polarized Gaussian beam[25] describes the transverse shift of the
beam as a whole, and the transverse shift is accord with the spin
Hall effect of photons in the geometrical optics. In next section,
the transverse shift will be reinvestigated, the interaction between
the polarization and the rotation deformation of a polarized
Gaussian beam will be presented besides the spin-orbit interaction.
\section{The correction of the spin Hall effect of a polarized Gaussian beam}
The complex-valued eikonal components can be written as
$S(\tau)=S^{R}(\tau)+iS^{I}(\tau)$, $S^{R}(\tau)$ and $S^{I}(\tau)$
are the real and imaginary components. Eq.(10) has the following
form
\begin{eqnarray}
exp[ik_{0}\psi(\textbf{r})]&=& exp[ik_{0}\psi_{c}]
exp\left[ik_{0}\left(\frac{\eta_{1}^{2}}{2}S_{11}+\frac{\eta_{2}^{2}}{2}S_{22}+\eta_{1}\eta_{2}S_{12}
+\eta_{1}S_{1}+\eta_{2}S_{2}\right)\right]\nonumber\\
&=&
exp[ik_{0}\psi_{c}]exp\left[ik_{0}\left(\frac{\eta_{1}^{2}}{2}S_{11}^{R}+\frac{\eta_{2}^{2}}{2}S_{22}^{R}+\eta_{1}\eta_{2}S_{12}^{R}
+\eta_{1}S_{1}^{R}+\eta_{2}S_{2}^{R}\right)\right]\nonumber \\
&
&exp\left[-k_{0}\left(\frac{\eta_{1}^{2}}{2}S_{11}^{I}+\frac{\eta_{2}^{2}}{2}S_{22}^{I}+\eta_{1}\eta_{2}S_{12}^{I}
+\eta_{1}S_{1}^{I}+\eta_{2}S_{2}^{I}\right)\right]\nonumber\\
&=& exp[ik_{0}\psi_{c}]\nonumber \\
& &
exp\left[ik_{0}\left(\frac{S_{11}^{R}}{2}(\eta_{1}+d_{1}^{R})^{2}+\frac{S_{22}^{R}}{2}(\eta_{2}+d_{2}^{R})^{2}
+\eta_{1}\eta_{2}S_{12}^{R}
-\frac{S_{11}^{R}}{2}(d_{1}^{R})^{2}-\frac{S_{22}^{R}}{2}(d_{2}^{R})^{2}\right)\right]\nonumber \\
&
&exp\left[-k_{0}\left(\frac{S_{11}^{I}}{2}(\eta_{1}+d_{1}^{I})^{2}+\frac{S_{22}^{I}}{2}(\eta_{2}+d_{2}^{I})^{2}
+\eta_{1}\eta_{2}S_{12}^{I}
-\frac{S_{11}^{I}}{2}(d_{1}^{I})^{2}-\frac{S_{22}^{I}}{2}(d_{2}^{I})^{2}\right)\right],
\end{eqnarray}
where $d_{1}$ and $d_{2}$ are
\begin{equation}
d_{i}^{R}=\frac{S_{i}^{R}}{S_{ii}^{R}},~~~~~~
d_{i}^{I}=\frac{S_{i}^{I}}{S_{ii}^{I}}.
\end{equation}
$d_{i}^{R}$ and $d_{i}^{I}$ are the transverse shift of the center
of the wave phase front and the cross section[33]. From Eq.(29), one
knows that the transverse shift[25] don't change the shape of the
phase front and the cross section of the beam, but the deformation
of the phase front and the cross section can affect the transverse
shift of the light beam. This transverse shift represents the spin
Hall effect of a polarized light beam[25].

By virtue of Eqs.(22-26) and Eq.(30), the evolution equations of the
transverse shift are obtained,
\begin{eqnarray}
\frac{\partial d_{1}^{I}}{\partial\tau}&=&
\frac{1}{S_{11}^{I}}\frac{\partial
S_{1}^{I}}{\partial\tau}-\frac{S_{1}^{I}}{(S_{11}^{I})^{2}}\frac{\partial
S_{11}^{I}}{\partial\tau}\nonumber\\
&=&
-\frac{\sigma}{2k_{0}}\frac{\partial\ln\varepsilon_{c}}{\partial\eta_{2}}
+\frac{\sigma}{2k_{0}}\frac{\partial\ln\varepsilon_{c}}{\partial\eta_{1}}\left(\frac{S_{12}^{I}}{S_{11}^{I}}\right)
+\mathcal {O}_{1}^{I},
\end{eqnarray}
where
\begin{equation}
\mathcal
{O}_{1}^{I}=\frac{1}{(S_{11}^{I})^{2}}[(S_{1}^{I}S_{11}^{R}-S_{1}^{R}S_{11}^{I})S_{11}^{I}
+(S_{1}^{I}S_{12}^{R}-S_{2}^{R}S_{11}^{I})S_{12}^{I}+(S_{1}^{I}S_{12}^{I}-S_{2}^{I}S_{11}^{I})S_{12}^{R}].
\end{equation}
Because of $d_{1}^{I}=S_{1}^{I}/S_{11}^{I}$, Eq.(31) is written as
\begin{eqnarray}
\frac{\partial d_{1}^{I}}{\partial\tau} &=&
-\frac{\sigma}{2k_{0}}\frac{\partial\ln\varepsilon_{c}}{\partial\eta_{2}}
+\frac{\sigma}{2k_{0}}\frac{\partial\ln\varepsilon_{c}}{\partial\eta_{1}}\left(\frac{S_{12}^{I}}{S_{1}^{I}}\right)d_{1}^{I}
+\mathcal {O}_{1}^{I}.
\end{eqnarray}
In the same way, we have
\begin{eqnarray}
\frac{\partial d_{2}^{I}}{\partial\tau} &=&
\frac{\sigma}{2k_{0}}\frac{\partial\ln\varepsilon_{c}}{\partial\eta_{1}}
-\frac{\sigma}{2k_{0}}\frac{\partial\ln\varepsilon_{c}}{\partial\eta_{2}}\left(\frac{S_{12}^{I}}{S_{2}^{I}}\right)d_{2}^{I}
+\mathcal {O}_{2}^{I},
\end{eqnarray}
\begin{eqnarray}
\frac{\partial d_{1}^{R}}{\partial\tau} &=&
-\frac{\sigma}{2k_{0}}\frac{\partial\ln\varepsilon_{c}}{\partial\eta_{2}}
+\frac{\sigma}{2k_{0}}\frac{\partial\ln\varepsilon_{c}}{\partial\eta_{1}}\left(\frac{S_{12}^{R}}{S_{1}^{R}}\right)d_{1}^{R}
+\mathcal {O}_{1}^{R},
\end{eqnarray}
\begin{eqnarray}
\frac{\partial d_{2}^{R}}{\partial\tau} &=&
\frac{\sigma}{2k_{0}}\frac{\partial\ln\varepsilon_{c}}{\partial\eta_{1}}
-\frac{\sigma}{2k_{0}}\frac{\partial\ln\varepsilon_{c}}{\partial\eta_{2}}\left(\frac{S_{12}^{R}}{S_{2}^{R}}\right)d_{2}^{R}
+\mathcal {O}_{2}^{R},
\end{eqnarray}
where
\begin{equation}
\mathcal
{O}_{2}^{I}=\frac{1}{(S_{22}^{I})^{2}}[(S_{2}^{I}S_{22}^{R}-S_{2}^{R}S_{22}^{I})S_{22}^{I}
+(S_{2}^{I}S_{12}^{R}-S_{1}^{R}S_{22}^{I})S_{12}^{I}+(S_{2}^{I}S_{12}^{I}-S_{1}^{I}S_{22}^{I})S_{12}^{R}],
\end{equation}
\begin{equation}
\mathcal
{O}_{1}^{R}=\frac{1}{(S_{11}^{R})^{2}}[(S_{1}^{I}S_{11}^{R}-S_{1}^{R}S_{11}^{I})S_{11}^{I}
-(S_{2}^{R}S_{12}^{R}-S_{2}^{I}S_{12}^{I})S_{11}^{R}+((S_{12}^{R})^{2}-(S_{12}^{I})^{2})S_{1}^{R}-S_{1}^{R}\alpha_{11}],
\end{equation}
\begin{equation}
\mathcal
{O}_{2}^{R}=\frac{1}{(S_{22}^{R})^{2}}[(S_{2}^{I}S_{22}^{R}-S_{2}^{R}S_{22}^{I})S_{22}^{I}
-(S_{2}^{R}S_{12}^{R}-S_{2}^{I}S_{12}^{I})S_{22}^{R}+((S_{12}^{R})^{2}-(S_{12}^{I})^{2})S_{2}^{R}-S_{2}^{R}\alpha_{22}].
\end{equation}
From Eqs.(33)-(36), the transverse shift in the plane
$(\eta_{1},\eta_{2})$ are
\begin{equation}
\textbf{d}^{R}=\frac{\sigma}{2k_{0}}\int(\textbf{l}\times\nabla\ln\varepsilon_{c})d\tau
+\frac{\sigma}{2k_{0}}\int\left(\frac{\partial\ln\varepsilon_{c}}{\partial\eta_{1}}\frac{\textbf{d}_{1}^{R}}{S_{1}^{R}}
-\frac{\partial\ln\varepsilon_{c}}{\partial\eta_{2}}\frac{\textbf{d}_{2}^{R}}{S_{2}^{R}}\right)S_{12}^{R}d\tau
+\mathcal {O}^{R},
\end{equation}
\begin{equation}
\textbf{d}^{I}=\frac{\sigma}{2k_{0}}\int(\textbf{l}\times\nabla\ln\varepsilon_{c})d\tau
+\frac{\sigma}{2k_{0}}\int\left(\frac{\partial\ln\varepsilon_{c}}{\partial\eta_{1}}\frac{\textbf{d}_{1}^{I}}{S_{1}^{I}}
-\frac{\partial\ln\varepsilon_{c}}{\partial\eta_{2}}\frac{\textbf{d}_{2}^{I}}{S_{2}^{I}}\right)S_{12}^{I}d\tau
+\mathcal {O}^{I},
\end{equation}
where $\mathcal {O}^{R}$ and $\mathcal {O}^{I}$ are
\begin{equation}
\mathbf{\mathcal {O}}^{R}=\int(\mathcal
{O}_{1}^{I}\textbf{e}_{1}+\mathcal {O}_{2}^{I}\textbf{e}_{2})d\tau,
\end{equation}
\begin{equation}
\mathbf{\mathcal {O}}^{I}=\int(\mathcal
{O}_{1}^{R}\textbf{e}_{1}+\mathcal {O}_{2}^{R}\textbf{e}_{2})d\tau.
\end{equation}
In Eq.(40) and Eq.(41),
$\textbf{d}=d_{1}\textbf{e}_{1}+d_{2}\textbf{e}_{2}$,$\textbf{d}_{i}=d_{i}\textbf{e}_{i}$.
From Eq.(40) and Eq.(41), the transverse shift of a polarized light
beam is different from photons. In the right-hand side of Eq.(40)
and Eq.(41), the first term corresponds to the transverse shift due
to the spin-orbit interaction, it describes the influence of the
interaction between the polarization and the trajectory of the
propagation of the light beam. The second term is closely related to
the quadratic terms $S_{12}$. The term $S_{12}$ describes the
rotation of the cross section of the beam in the plane
$(\eta_{1},\eta_{2})$[23,30,31]. The rotation of the beam produces
an additional equivalent intrinsic orbital angular
momentum[26,27,28], so the second term corresponds to the
spin-intrinsic orbital angular momentum interaction[34]. The third
term describes the influence of the deformation on the transverse
shift. One has known that the propagation of a polarized light(or
photons) is influenced by the spin-orbit interaction, but a
polarized light beam is different from photons because of the
spatial energy distribution of the light beam. This result means
that it is not enough if we only consider the influence of the
spin-orbit interaction for the propagation of a polarized light beam
in an inhomogeneous isotropic medium, the rotation deformation of
the beam also is important. This is the main result in this paper.

The imaginary components of $S_{ij}$ describe the amplitude(energy)
spatial distribution of the Gaussian beam. From Eq.(40) and Eq.(41),
the splitting of the center of gravity for the linear polarized
Gaussian beam is obtained
\begin{equation}
\delta\textbf{d}^{I}=(\textbf{d}^{I})^{+}-(\textbf{d}^{I})^{-}=\frac{1}{k_{0}}\int(\textbf{l}\times\nabla\ln\varepsilon_{c})d\tau
+\frac{1}{k_{0}}\int\left(\frac{\partial\ln\varepsilon_{c}}{\partial\eta_{1}}\frac{\textbf{d}_{1}^{I}}{S_{1}^{I}}
-\frac{\partial\ln\varepsilon_{c}}{\partial\eta_{2}}\frac{\textbf{d}_{2}^{I}}{S_{2}^{I}}\right)S_{12}^{I}d\tau.
\end{equation}
Here, the first term describes the splitting due to the spin-orbit
interaction, the second term corresponds to the splitting due to the
rotation deformation of the Gaussian beam. This is a significant
difference between the light beam and photons.

Although the analytic solutions of the differential equations
(22)-(27) cannot be obtained, we give the numerical simulation when
the linear polarized Gaussian beam propagates along the helical ray
in the cylindrical symmetry medium[23]. Fig.1 shows the evolution of
the splitting for the center(central ray) of the linear polarized
Gaussian beam and the propagation trajectory of photons. Because of
the deformation of a polarized Gaussian beam, the relative shift
between the right-hand and left-hand circularly polarized Gaussian
beam is slightly different from photons. Although the difference of
the relative shift is very small, one also should pay attention to
the spin-intrinsic orbital angular momentum interaction because of
the deformation of a polarized light beam.

\section{Conclusion}
The propagation of a polarized Gaussian beam in an inhomogeneous
isotropic medium has been presented using the description of the
eikonal-based complex geometrical optics in this paper. The linear
complex-valued terms are introduced to describe the perturbation of
the center of a polarized Gaussian beam due to the inhomogeneity of
medium and the deformation of a polarized light  beam. Because the
rotation deformation of a polarized Gaussian beam produces an
additional equivalent intrinsic orbital angular momentum, the
interaction between the polarization and the rotation deformation of
a polarized light beam is presented. This interaction can be
considered as the spin-intrinsic orbital angular momentum and makes
the correction for the spin Hall effect of a polarized light beam.

\newpage
\section*{References}
\label{}



\begin{thebibliography}{48}




\bibitem[Liberman et al.(1992)]{Liberman92}
V S Liberman, B Ya Ze$l^{,}$dovich, Phys. Rev. A, \textbf{46}, 5199
(1992).

\bibitem[Onada et al.(2004)]{Onada04}
M Onada, S Marakami, N Nagaosa , Phys. Rev. Lett, \textbf{93},
083901 (2004).

\bibitem[Bliokh et al.(2004)]{Bliokh04}
K Yu Bliokh, Yu P Bliokh ,Phys. Rev. E, \textbf{70}, 026605 (2004).

\bibitem[Bliokh et al.(2005)]{Bliokh05}
K Yu Bliokh, V D Freilikher ,Phys. Rev. B,
\textbf{72}, 035108 (2005).

\bibitem[Bliokh et al.(2006)]{Bliokh06}
K Yu Bliokh, Yu P Bliokh ,Phys. Rev. Lett, \textbf{96}, 073903
(2006).

\bibitem[Duval et al.(2006)]{Duval06}
C Duval, Z Horv$\acute{a}$th, P A Horv$\acute{a}$thy,Phys. Rev. D,
\textbf{74}, 021701 (2006).

\bibitem[Berard et al.(2006)]{Berard06}
A B$\acute{e}$rard , H Mohrbach, Phys.Lett.A, \textbf{352}, 190
(2006).

\bibitem[Gosselin et al.(2007)]{Gosselin07}
P Gosselin, A B$\acute{e}$rard , H Mohrbach, Phys. Rev. D,
\textbf{75}, 084035 (2007).

\bibitem[Bliokh et al.(2007)]{Bliokh07}
K Yu Bliokh, D Yu Frolov, Y A Kravtsov ,Phys. Rev. A, \textbf{75},
053821 (2007).

\bibitem[Bliokh et al.(2008)]{Bliokh08}
K Yu Bliokh ,A Niv , V Kleiner, E Hassman ,Nature photonics
\textbf{2}, 748 (2008).

\bibitem[Hosten et al.(2008)]{Hosten08}
O Hosten, P Kwiat, Science, \textbf{319}, 787 (2008)

\bibitem[Hailu et al.(2009)]{Hailu09}
Hailu Luo, Shuangchun Wen, Weixing Shu, Zhixiang Tang, Yanhong Zou,
Dianyuan Fan, Phys. Rev. A, \textbf{80}, 043810 (2009).

\bibitem[Bliokh et al.(2010)]{Bliokh10}
K Yu Bliokh, M A Alonso, E A Ostrovskaya, A Aiello, Phys. Rev. A,
\textbf{82}, 063825 (2010).

\bibitem[Hailei et al.(2011)]{Hailei11}
Hailei Wang, Xiangdong Zhang, Phys. Rev. A, \textbf{83}, 053820
(2011).

\bibitem[Guoding et al.(2011)]{Guoding11}
Guoding Xu, Taocheng Zang, Hongmin Mao, Tao Pan, Phys. Rev. A,
\textbf{83}, 053828 (2011).

\bibitem[Shitrit et al.(2011)]{Shitrit11}
N Shitrit, I Bretner, Y Gorodetski, V Kleiner, E Hasman, Nano
Letters, \textbf{11}, 2038 (2011).

\bibitem[Berry et al.(1984)]{Berry84}
M V Berry, Proc. Roy. Soc. A, \textbf{392}, 45 (1984).

\bibitem[Rytov et al.(1938)]{Rytov38}
S M Rytov, Dokl. Akad. Nauk. SSSR, \textbf{18}, 263 (1938).

\bibitem[Chiao et al.(1986)]{Chiao86}
R Y  Chiao and Y S Wu, Phys. Rev. Lett. \textbf{57}, 933 (1986).

\bibitem[Tomita et al.(1986)]{Tomita86}
A Tomita and R Y Chiao, Phys. Rev. Lett. \textbf{57}, 937 (1986).

\bibitem[Kravtsov et al.(2004)]{Kravtsov04}
Y A Kravtsov ,Geometrical Optics in Engineering Physics, Alpha
Science,2004.

\bibitem[Berczynski et al.(2004)]{Berczynski04}
P Berczynski , Y A Kravtsov ,Phys.Lett.A, \textbf{331}, 265 (2004).

\bibitem[Berczynski et al.(2006)]{Berczynski06}
P Berczynski , K Y Bliokh ,Y A Kravtsov , A Stateczny,
J.Opt.Soc.Am.A, \textbf{23}, 1442 (2006).

\bibitem[Bornatici et al.(2006)]{Bornatici06}
M Bornatici , O Maj , Plasma Phys.Controlled Fusion, \textbf{45},
707 (2003).

\bibitem[Bliokh et al.(2009)]{Bliokh09}
K Y Bliokh , A Desyatnikov, Phys. Rev. A, \textbf{79}, 011807
(2009).

\bibitem[Alexeyev et al.(2005)]{Alexeyev05}
C N Alexeyev ,M A Yavorsky , J.Opt.A: Pure Appl.Opt. \textbf{7}, 416
(2005).

\bibitem[Bekshaev et al.(2006)]{Bekshaev06}
A Bekshaev , Opt. Lett, \textbf{31}, 2199 (2006).

\bibitem[Kotlyar et al.(2007)]{Kotlyar07}
V V Kotlyar, S N Khonina, R V Skidanov, V A Soifer, Opt. Com,
\textbf{274}, 8 (2007).

\bibitem[Kravtsov et al.(1990)]{Kravtsov90}
Y A Kravtsov , Y I Orlov, Geometrical Optics of Inhomogeneous
medium, Springer-Verlag,1990.

\bibitem[Permitin et al.(1996)]{Permitin96}
G V Permitin , A I Smirnov , J.Exp.Theor.Phys, \textbf{82}, 395
(1996).

\bibitem[Permitin et al.(2001)]{Permitin01}
G V Permitin , A I Smirnov , J.Exp.Theor.Phys, \textbf{92}, 10
(2001).

\bibitem[Born et al.(1999)]{Born99}
M Born ,E Wolf ,7th edt Principles of Optics,Cambridge University
Press, 1999

\bibitem[Casperson et al.(1973)]{Casperson73}
L W Casperson ,Appl.Opt \textbf{12}, 2434 (1973).

\bibitem[Allen et al.(1996)]{Allen96}
L Allen ,V E Lembessis, M Babiker, Phys. Rev. A, \textbf{53}, 2937
(1996).

\end{thebibliography}



\newpage

\begin{figure}
\centering
\begin{minipage}[b]{1\textwidth}
\centering
\includegraphics[width=3in]{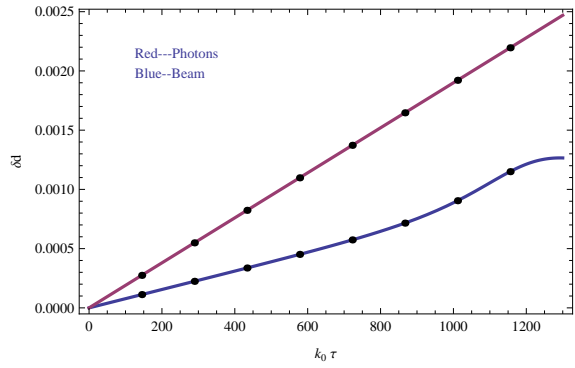}
\end{minipage}
\caption{Numerical result for the splitting of the linear polarized
Gaussian beam and photons which both propagate along the helical ray
in the cylindrical symmetrical inhomogeneous medium. The red line
and the blue line describe the evolution of the splitting of the
trajectory of the beam and photons respectively. The parameters are
as follows: the angle between the tangent to the helical ray and the
cylinder axis, the dielectric permittivity
$\varepsilon=\varepsilon_{0}-\rho^{2}/L^{2}$, $\varepsilon_{0}=1$,
$L=200\lambda$, the initial beam waist radius $w_{0}=10\lambda$, the
initial phase front is flat $r=\infty$.}
\end{figure}

\end{document}